\documentstyle[12pt]{article}
\begin{document}

\begin{center}

\large{{\bf In Quest of a True Model of the Universe\footnote{Received Honorable Mention in the essay competition 2004, of the Gravity Research Foundation} }}
\end{center}

\medskip
\begin{center}
R. G. Vishwakarma\\ 

\vspace{0.5cm}
\emph{Department of Mathematics\\
    Autonomous University of Zacatecas\\
  Zacatecas, ZAC C.P. 98060\\
                Mexico}\\
Email: rvishwa@mate.reduaz.mx
\end{center}

\medskip
\begin{abstract}\noindent
While many observations support the validity of Einstein's general relativity
as the theory of gravity, there are yet many that suggest the presence of new
physics.
In order to explain the high-redshift supernovae Ia 
observations together with the recently made precise observations of the CMB 
anisotropy by WMAP, the standard cosmology  has to invoke some 
hypothetical matter with unnatural properties which is very speculative. 
This casts doubts upon the foundations of the standard cosmology and suggests
that some theoretical concept may still be missing from the theory.

Such a concept might be the rotation of the astronomical objects, which 
has not been properly taken care of when we claim that a perfect fluid is 
a good approximation to the real contents of the universe.
A crude estimation of the angular kinetic energy of massive 
galaxies indicates to a possibility to have $\Omega_{\rm total}\approx 1$ without invoking the 
hypothetical dark matter or dark energy. This picture also appears consistent with the 
recent observations of a great abundance of old massive galaxies made by Gemini 
Deep Deep Survey. However, a proper relativistic theory of the rotating objects is still to be
investigated. It is expected that the consequences of 
incorporating rotation in general relativity, and hence in special relativity, would be profound.

\end{abstract}

\newpage
\noindent
The present-day cosmology is passing through a critical phase. If one believes 
in the standard big bang cosmology, then the current observations seem 
to favour
a universe in which most of the content is `dark', which though does not
appear to resemble any known form of matter tested in the laboratory. This 
hypothetical constituent is required by the theory to occur in two forms:

\medskip
\noindent
(i) First is the dark matter, consisted of non-baryonic  
particles originally predicted to solve the problems of structure formation and of the 
missing mass in bound gravitational systems such as galaxies and clusters of 
galaxies. 
However, there is not only disagreement over whether dark matter is predominantly cold or 
hot, but also there is no satisfactory observational evidence for the postulated particles in
laboratory physics. 
The 
predicted density distribution of dark halos which result from N-body 
simulations \cite{nbody}, appears to 
be inconsistent with a number of observations on the galactic and smaller scales \cite{sg} 
or with strong lensing in clusters of galaxies \cite{gl}.
The most favoured candidate of dark matter postulated by many astrophysicists, cosmologists 
and particle physicists is a massive but very weakly interacting particle called WIMP.
Currently there are a number of WIMP detection experiments underway. Among these, the
DAMA experiment \cite{dama}, which measures the annual modulation in WIMP interactions 
with the sodium-iodide detectors caused by the earth's rotation around the sun, is the 
only one to have claimed a positive signal. However, the results of this experiment are 
controversial as other more sensitive searches have not detected nuclear recoils due to 
WIMP interactions \cite{cdms} and concluded that almost all the events measured by DAMA
were from neutrons, and should not be attributed to scattering events from dark-matter 
WIMPs. 
It is therefore fair to say that this scheme has still to demonstrate its viability.

\medskip
\noindent
(ii) Second is the dark energy having a negative pressure, which is even more mysterious. This may be identified with the cosmological constant $\Lambda$ or the 
zero-point energy density of vacuum. This poses the well known formidable
problems which have not been solved yet: Why does it have an unnaturally low energy density, 
$10^{-122}$ in Planck units? And if the energy density of vacuum remains constant
but that of matter dilutes with the expansion of the universe, why are their energy
densities so comparable at the present epoch? A popular approach to these problems is 
to invoke an ad-hoc self-interacting potential (quintessence field). However, firstly, the
quintessence field has not been identified, and secondly, a derivation of the potential 
from the first principles is still lacking.

Apart from this, there are also other discrepancies between the theory and the 
observations. We mention the following two.

\noindent
(i) The distances measured in cosmology by two different methods $-$ the luminosity 
distance $d_{\rm L}$ (by measuring the apparent luminosity of standard candles) and the angular 
diameter distance  $d_{\rm A}$ (by measuring the apparent size of the standard rulers)
$-$ are related
by
\begin{equation}
d_{\rm L} (z) = d_{\rm A} (z) (1+z)^2,
\end{equation} 
which is known as reciprocity or distance duality relation and is expected to hold 
for general metric theories of gravity. However, the combined supernovae (SNe) Ia 
data \cite{sn}
(providing $d_{\rm L}$) and the data on the radio galaxies, compact radio sources and X-ray 
clusters \cite{duality}
(providing $d_{\rm A}$) are in disagreement with the reciprocity relation at $2\sigma$
\cite{duality}.

\noindent
(ii)
There is also a discrepancy between the light elements
abundances and the Wilkinson Microwave Anisotropy Probe (WMAP) \cite{cmb}. 
The baryon density measured by the WMAP experiment
or the primordial D abundance is much higher than the one measured by the 
He$^4$ or Li$^7$ abundances \cite{ichikawa}.

\medskip
Most works in cosmology are, at present, dedicated to refining minor details of the 
standard cosmology and do not care about having a deeper insight of the 
foundations of the theory. The possibility is that some theoretical concept might
still be missing from the theory. In this essay, we point out towards such a concept
which, as far as we know, has not been taken care of in the framework of the standard 
cosmology. This is the rotation of the astronomical objects. Certainly this 
has not been properly taken care of when we claim that a perfect fluid is 
a good approximation to the real contents of the universe.

We know that the largest structures of the universe,
the galaxies, are rotating. These contain subsystems, from solar-like systems to
stars to planets to asteroids, which all rotate. They all possess (kinetic) energy of 
rotation associated with their angular momenta. As all kinds of energy and momenta couple
to gravity (according to Einstein´s theory), this energy must also find space in the Einstein 
field equations. One may, however, argue that the angular momenta of different galaxies, oriented randomly, 
will cancel out on large scales. However, this cannot be true for the energy
associated with their rotations, which is a scalar quantity. In the following, we shall 
see that if this effect is taken into account, in the way the recent observations indicate,
it will provide significant contribution to the cosmic energy density and there 
is a possibility to  explain the observations without invoking
the hypothetical dark matter and dark energy.

In order to incorporate the rotation of galaxies into Einstein´s field equations,
the correct way would be to find the energy-momentum-angular momentum tensor of the
matter content, which though requires more theoretical understanding of the problem
and its consequences for related issues. However, one can make a very crude estimate
of the contribution to the energy density of the universe from this effect in the following way.
If one approximates the milky way by a rigidly rotating flat disk with constant density, then
the kinetic energy associated with its rotation can be estimated by
\begin{equation}
\mbox{angular KE} = \frac{1}{4}M R^2 \omega^2,
\end{equation} 
where $M\approx (7.5-10) \times 10^{11} M_{\odot}$ is the mass of the milky way and $R$ and 
$\omega$ are its radius and angular speed respectively. By considering
$R\approx$ 25 kpc and $\omega\approx$ 1 revolution per
220 million years, this energy comes out as $\approx 10^{11} M$ Joules
(where $M$ is measured in kg). This is equivalent to a mass ($m=E/c^2$) of
$10^{-6}M$ kg. 
Thus the ratio of this mass to the mass of the milky way, or the ratio of the corresponding 
energies, is about $10^{-6}$. This is also equal to
the ratio of the corresponding energy densities (obtained by dividing this ratio by the average 
volume of space per galaxy). In order to compare the contribution of the angular KE density with
the cosmological energy density, one can write the ratio in terms of the energy density parameter
$\Omega \equiv \rho/\rho_{\rm critical}$ (i.e., the energy density in units of the critical density
 $\rho_{\rm critical}=3H^2/8\pi G$).
Thus the contribution of the angular kinetic energy density to the cosmological energy density
from the milky way-like galaxies
is negligible: $\Omega_{\rm rotation}\approx 10^{-6} ~\Omega_{\rm matter}$ 
only. By modelling the galaxies by ellipsoidal or spherical balls, one gets
a similar order of contribution.
The corresponding contribution from the individual stars is even smaller.
However, for more massive galaxies this contribution can be significantly higher.
Note that our own milky way is on the light weight-side.
More massive galaxies rotate faster, because the centrifugal force balances out the increased 
gravity (here we are talking about the average angular speed of the galaxy as a whole). 
Thus for a galaxy with a diameter only 25 times that of the milky way and angular speed only 25 
times that of the milky 
way, the $\Omega_{\rm rotation}$ will be of the order of its  $\Omega_{\rm matter}$. (The numbers 
chosen are just random whole numbers and are not indicative of any scaling between size and mass of 
the galaxies. Any such scaling can be given by more accurate observations of
galaxies.) 
For a galaxy with size and angular speed only as large as 50 times those of the milky 
way, the $\Omega_{\rm rotation}\approx 10 ~ \Omega_{\rm matter}$. And for a galaxy with 
size and angular speed as large as 100 times those of the milky 
way, the $\Omega_{\rm rotation} \approx 150 ~ \Omega_{\rm matter}$. 
Although these contributions come from a crude modelling of the galaxies, which have a rather
more complicated dynamics in reality, however the contributions are too large to be neglected. 
We also admit that the estimates of the angular kinetic energy 
obtained above, which are of the order of the rest mass of the objects, should not  be treated in a non-relativistic way. However, for more accurate 
estimates, we have to wait until a full relativistic theory is investigated which incorporates the rotation of galaxies in a natural way.

Recalling that  
$\Omega_{\rm matter}\approx 0.04$, it seems that it is very much possible to have  
$\Omega_{\rm total}\approx 1$, without invoking
any hypothetical dark matter or energy, if one just takes account of the energy associated with the 
rotation of massive galaxies and if the universe is laden with massive galaxies. This 
picture also appears consistent with the recently made direct observations of a great abundance
of massive old galaxies by Gemini Deep Deep Survey (GDDS) \cite{gemini}.
This survey, which used a special technique to capture even the faintest galactic light, 
covered more area than the previous surveys and found a great abundance of giant galaxies with
mass $>10^{10.8}M_{\odot}$ at $z>1$. These observations pose problem for the current hierarchical 
paradigm of galaxy formation which believes that massive galaxies form from the merger of smaller 
units. This process is however slow and takes billions of years. However the survey
found these fully formed and mature galaxies at a time when the universe was just about 20$-$40 
percent of its present age.
This  raises the question ``why do the galaxies in the young universe appear so mature?´´

One may doubt whether this way of making $\Omega_{\rm total}\approx 1$ without invoking 
any dark energy would be consistent with the current observations.     
There seems to be an impression in the community that the current observations,
particularly the high redshift SNe Ia observations and the measurements of the angular
power fluctuations of the CMB, can be explained only in the framework of
an accelerating universe. This, however, does not seem correct.
Although the interpretation of the CMB anisotropies has emerged as the single
most important tool to examine a theory, one must bear in mind that the
conclusion drawn do rest upon a number of assumptions, and the results are
altogether not as robust as we are often led to believe. 
Taken on their face value, the CMB observations made by WMAP make the only 
apparent prediction that $\Omega_{\rm total}\approx 1$
and the decelerating models (for example, the CDM Einstein-de Sitter model: 
$\Lambda=0$, $\Omega_{\rm total}=1$) are also not ruled out 
\cite{vishwa1}.
The accelerating expansion is mainly motivated by the high redshift supernovae Ia observations.
It however seems that the predictions of these observations depend on how they are analysed, 
which though is not correct. When the low- and the high-redshift data points are analysed 
separately, they admit decelerating models without having any dark energy.
However, when combined together, they rule out these models \cite{paddy}. Ironically
the updated SNe Ia data set even rules out the {\it preferred} flat model with $\Lambda$ at a high
degree of significance, which was favoured by the earlier version of the data \cite{paddy}.
This casts doubts upon the credibility of these observations and until more accurate
data is available, one cannot  believe their predictions.
It may also be noted that in the earlier version of data (which rules out the decelerating models 
otherwise),
if one takes into account the absorption of light by the inter-galactic metallic dust
(ejected from the supernovae explosions)
which extinguishes radiation travelling over long distances, then
the observed faintness of the extra-galactic SNe Ia can be explained successfully
in the decelerating models \cite{vishwa1},\cite{vishwa2}.

If one however wants to impose the requirement for an accelerating expansion in the present
framework of rotation of galaxies, then this would imply 
that the galaxies do not have random orientations; so that their angular
momenta do not cancel out completely. In this case there will be a net,
non-zero angular momentum left out, providing a global rotation to the 
universe. It has been shown that a rotating universe produces an
accelerating expansion \cite{godlowski}.
%
%
This would also mean that the rotation axes of
the galaxies are not altogether randomly oriented. This might also explain 
why galaxies have the
shape and size they have and why most of the universe is dominated by structures 
that violate radial symmetry.

In this letter, we have tried to drag attention towards a concept, the rotation of galaxies, which
has  not been
properly taken care of when we claim that a perfect fluid is a good
approximation to the real contents of the universe.
Our crude estimate of the 
contribution to the cosmic energy density from the angular kinetic energy of massive galaxies, 
observed by GDDS, appears to be too large to be 
neglected and possibly can close the universe even without invoking dark matter and dark energy.
However, a proper relativistic theory of rotating astronomical objects is still missing. The consequences of 
incorporating rotation in general relativity (and hence also in special relativity), would be profound \cite{inprogress}.
This might prove instrumental in explaining different kinds of rotation curves of galaxies (without invoking dark matter) as the resulting theory would possess more dynamical details of galaxies than the existing theory. It may be noted that in addition to the nearly flat rotation curves and slightly rising 
rotation curves of spiral galaxies \cite{rising}, there have also been observed spiral galaxies
with sharply declining rotation curves \cite{declining}. The usual explanations, based on Newtonian
theory of gravitation, invoke different assumptions for individual galaxies. 
In this context, the models based on the Modified Newtonian Dynamics 
(MOND) are noteworthy, which (they however do not require any dark matter)  have met some success
\cite{ndm} and it is expected that a general covariant theory (MOND is not a 
covariant theory) should reduce to a MOND-like physics in the galaxy formation era. 

At the moment, it is not clear how the theory we propose would reduce to a 
MOND-like physics in a particular era. It is also not guaranteed that one would be
able to make $\Omega_{\rm total}=1$ from this theory alone (for which more accurate observations of the details of galactic dynamics are needed). However, a
concept which has not been taken care of in the existing theories of gravity, and which seems to yield important non-negligible implications, must be taken
seriously. We should try to investigate a proper relativistic theory for
rotating galaxies. 

\vspace{0.5cm}

\noindent
{\bf Acknowledgement}

The author thanks Henning Knutsen and Jayant V. Narlikar for useful comments and discussion. Thanks are also due to the Abdus Salam ICTP for sending the necessary literature 
whenever required under the author's associateship programme.


\vspace{0.5cm}

\noindent
{\bf References}

\begin{enumerate}



\bibitem{nbody} Navarro, J. F. et al, (1996) Astrophys. J., {\bf 462}, 563.
 
\bibitem{sg} Ghigna, S. et al, (2000) Astrophys. J., {\bf 544}, 616;
             de Blok, W. J. G. et al, (2001) Astron. J., {\bf 122}, 2396.

\bibitem{gl} Treu, T. et al, (2003) astro-ph/0311052.

\bibitem{dama} Bernabei, R. et al, (2003) Riv. Nuovo Cim., {\bf 26}, 1.

\bibitem{cdms} Akerib, D. S. et al, (2004) astro-ph/0405033;
               Angloher, G. et al, (2004) astro-ph/0408006.

\bibitem{sn} Barris, B. J. et al., (2004) Astrophys. J. {\bf 602}, 571;
             Tonry, J. L.et al., (2003) Astrophys. J. {\bf 594}, 1;
             Knop, R. A. et al., (2003) astro-ph/0309368 (to appear in Astrophys. J.).

\bibitem{duality} Bassett, B. A., and Kunz, M., (2003) astro-ph/0312443.

\bibitem{cmb} Page, L. et al., (2003) Astrophys. J. Suppl. {\bf 148}, 233.

\bibitem{ichikawa} Ichikawa, K., Kawasaki, M. and Takahashi, F., (2004) astro-ph/0402522.

\bibitem{gemini} Glazebrook, K., et al., (2004) astro-ph/0401037;
                Abraham, R. G., et al., (2004) astro-ph/0402436 (to appear in Astron. 
J.).

\bibitem{vishwa1} Vishwakarma, R. G., (2003) Mon. Not. Roy. Astron. Soc., 
                  {\bf 345}, 545.

\bibitem{paddy} RoyChoudhury, T. and Padmanabhan, T., (2003) astro-ph/0311622
                 (to appear in Astron. Astrophys.).

\bibitem{vishwa2} Vishwakarma, R. G., (2002) Mon. Not. Roy. Astron. Soc., 
                         {\bf 331}, 776.

\bibitem{godlowski}  Godlowski, W. and Szydlowski, M. (2003) Gen. Rel. Grav. 
                         {\bf 35}, 2171;
                   Godlowski, W. and Szydlowski, M. (2004) astro-ph/0409073.

\bibitem{inprogress} Vishwakarma, R. G. (in progress).

\bibitem{rising} Rubin, V. C., (1997) ``Bright Galaxies, Dark Matter", AIP Press, 
                   American Institute of Physics.

\bibitem{declining} Carignan, C. and Puche, D., (1990) Astron. J. {\bf 100}, 394;
                    Casertano, S. and van Gorkom, J. H., (1991) Astron. J. {\bf 101}, 1231.

\bibitem{ndm} Milgrom, M., (1983) Astrophys. J. {\bf 270}, 371;
              Begeman, K. G., Broeils, A. H. and Sanders, R. H., (1991) 
                  Mon. Not. Roy. Astron. Soc., {\bf 249}, 523;
              Sanders, R. H., (1996) Astrophys. J.,{\bf 473}, 117;
              Sanders, R. H. and McGaugh, S. S., (2002) Ann. Rev. Astron. 
                  Astrophys., {\bf 40}, 263;
              Milgrom, M. and Sanders, R. H., (2003) Astrophys. J.


\end{enumerate}
\end{document}